\documentclass[12pt]{iopart}
\usepackage{times}
\usepackage{graphicx}

\newcommand{\lesssim}{\raisebox{0.5mm}{\em $\, <$} \hspace{-3.3mm}
\raisebox{-1.5mm}{\em $\sim \,$}}
\newcommand{\gtrsim}{\raisebox{0.3mm}{\em $\, >$} \hspace{-3.3mm}
\raisebox{-1.5mm}{\em $\sim \,$}}
\def\mbb{\langle m \rangle_{\beta\beta}}
\def\ma{{\mbox{\scriptsize{max}}}}
\def\mi{{\mbox{\scriptsize{min}}}}
\def\CH{{\mbox{\scriptsize{CH}}}}
\def\eV{\mbox{eV}}

\begin{document}

\title{Constraints on Neutrino Parameters
by Neutrinoless Double Beta Decay Experiments}
\author{Hiroaki Sugiyama
\footnote{
Invited talk based on \cite{paper}
at Beyond the Desert 02, Oulu, Finland, 2-7 June 2002.}
}
\address{Department of Physics, Tokyo Metropolitan University\\
1-1 Minami-Osawa, Hachioji, Tokyo 192-0397, Japan\\
E-mail: hiroaki@phys.metro-u.ac.jp}

\begin{abstract}
 Allowed regions on the $m_l$ - $\cos{2\theta_{12}}$ plane are extracted
from results of neutrinoless double beta decay experiments.
 It is shown that
$0.05\,\eV (0\,\eV) \lesssim m_l \lesssim 1.85\,\eV$ is obtained
for the normal (inverted) hierarchy
by using the LMA best fit parameters and
the $0\nu\beta\beta$ result announced late last year,
which is $0.05\,\eV \le \mbb \le 0.84\,\eV$
with $\pm 50\,\%$ uncertainty of the nuclear matrix elements.
\end{abstract}

\section{Introduction}
 Although it is well known that neutrinos are massive
by neutrino oscillation experiments~\mbox{[2-4]},
the values of their masses are still unknown.
 The answer is never given by oscillation experiments
because oscillation probabilities depend on
$\Delta m^2_{ij} \equiv m^2_j - m^2_i$.
 Thus, we rely on non oscillation experiments
such as single beta decay measurements~\cite{beta},
which are direct measurements of neutrino mass,
neutrinoless double beta decay ($0\nu\beta\beta$) searches~\cite{betabeta}
and cosmological measurements~\cite{cosmological}
or Z-burst interpretation of the highest energy cosmic ray~\cite{Zburst}.

 Double beta decay experiments seems to have
rather higher sensitivities than those of other non oscillation experiments.
 Recent results of negative observations of $0\nu\beta\beta$
put upper bounds on the observable $\mbb$
as $\mbb < 0.35\,\eV$ \mbox{(90\,\% C.L.)}
by Heidelberg-Moscow~\cite{HeidelMoscow}
and $\mbb < 0.33-1.35\,\eV$ \mbox{(90\,\% C.L.)} by IGEX~\cite{IGEX}
\footnote{
 The bounds on $\mbb$ depend on nuclear matrix elements.
 The actual observable is the half-life $T_{1/2}^{0\nu}$:
 $T_{1/2}^{0\nu} > 1.9\times10^{25}$~y was obtained by Heidelberg-Moscow,
and $T_{1/2}^{0\nu} > 1.57\times10^{25}$~y by IGEX.}.
 The energy regions to be probed can reach to
the order of $10^{-2}\,\eV$ by some of the future experiments [11-17].
 Such experiments seem to have strong possibility of
$0\nu\beta\beta$ observations.
 Actually,
an observation of $0\nu\beta\beta$ was announced late last year
as $0.05\,\eV \le \mbb \le 0.84\,\eV$ \mbox{(95\,\% C.L.)}
with $\pm 50\,\%$ uncertainty of the nuclear matrix elements
(KDHK result~\cite{evidence};
 See also comment on the results and the replies~\cite{comment}).
 While this result should be checked in the future experiments,
it is fruitful to investigate what kind of information
can be extracted from $0\nu\beta\beta$ observations.

 The constraints imposed on neutrino mixing parameters
by $0\nu\beta\beta$ experiments
have been discussed by many authors (See the references in~\cite{paper}).
 The implications of the KDHK result have been also discussed
(See, for example, the references of the second article in~\cite{comment}).
 In this talk,
the constraints on a neutrino mass and the solar mixing angle
are discussed in the generic three flavor mixing framework
by observations
(as well as non-observations) of $0\nu\beta\beta$.

\section{Constraints}
 We use the following standard parametrization
of the MNS matrix~\cite{MNS}:
\begin{equation}
 U_{\mbox{\scriptsize{MNS}}}
  \equiv
  \left[
   \begin{array}{ccc}
    c_{12}c_{13} & s_{12}c_{13} & s_{13}e^{-i\delta}\nonumber\\
    -s_{12}c_{23}-c_{12}s_{23}s_{13}e^{i\delta} &
     c_{12}c_{23}-s_{12}s_{23}s_{13}e^{i\delta} & s_{23}c_{13}\nonumber\\
    s_{12}s_{23}-c_{12}c_{23}s_{13}e^{i\delta} &
     -c_{12}s_{23}-s_{12}c_{23}s_{13}e^{i\delta} & c_{23}c_{13}\nonumber\\
   \end{array}
  \right].
\label{MNSmatrix}
\end{equation}
 The mixing matrix for three Majorana neutrinos is
\begin{equation}
 U \equiv U_{\mbox{\scriptsize{MNS}}}
           \times \mbox{diag}(1, e^{i\beta}, e^{i\gamma}) ,
\end{equation}
where $\beta$ and $\gamma$ are extra CP-violating phases
which are characteristic of Majorana particles~\cite{Mphase}.
 In this parametrization, the observable of double beta decay experiments
is described as
\begin{eqnarray}
 \mbb
 &\equiv& 
  \left|\, \sum^{3}_{i=1} m_i U^2_{ei}\, \right| \nonumber\\
 &=&
  \left|\,
   m_1 c_{12}^2 c_{13}^2
   + m_2 s_{12}^2 c_{13}^2 e^{2 i \beta}
   + m_3 s_{13}^2 e^{2 i (\gamma - \delta)}\,
  \right| ,
\label{beta1}
\end{eqnarray}
where $U_{ei}$ denote the elements in the first low of $U$
and $m_i$ ($i=1,2,3$) are the neutrino mass eigenvalues.
 In the convention of this talk,
the normal hierarchy means $m_1 < m_2 < m_3$
and the inverted hierarchy $m_3 < m_1 < m_2$.

 In order to utilize $\mbb^\ma$
which is an experimental upper bound on $\mbb$, 
we derive a theoretical lower bound on $\mbb$.
 An appropriate choice of the phase-factor
$e^{2 i(\gamma - \delta)}$ in eq.~(\ref{beta1})
leads to an inequality
\begin{eqnarray}
 \mbb^\ma
 \ge \mbb
 \ge c_{13}^2
         \left|
	  m_1 c_{12}^2 + m_2 s_{12}^2 e^{2 i \beta}
         \right|
       - m_3 s_{13}^2 .
\label{beta2}
\end{eqnarray}
 Strictly,
the right-hand side (RHS) should be the absolute value of it.
 It is, however, not necessary to consider the absolute value
because $s_{13}^2$ has a very small value.
 The RHS of (\ref{beta2}) is minimized
by replacing $e^{2 i \beta}$ with $-1$
and $s_{13}^2$ with the largest value $s_\CH^2$ ($\simeq 0.03$)
which is determined by reactor experiments~\cite{CHOOZ}.
 Thus, we obtain
\begin{eqnarray}
 \mbb^\ma
 \ge c_\CH^2 \left| m_1 c_{12}^2 - m_2 s_{12}^2 \right|
      - m_3 s_\CH^2 .
 \label{beta3}
\end{eqnarray}

 Next,
we derive a theoretical upper bound on $\mbb$
to utilize an experimental lower bound $\mbb^\mi$.
 Since the RHS of (\ref{beta1})
is maximized by setting the phase-factors unity,
we obtain
\begin{eqnarray}
 \mbb^\mi
  \le \left(m_1 c_{12}^2 + m_2 s_{12}^2 \right) c_{13}^2
        + m_3 s_{13}^2 .
\label{beta4}
\end{eqnarray}
 Furthermore,
$s_{13}^2$ is replaced by $s_\CH^2$ (zero)
for the normal (inverted) hierarchy
in order to set the RHS to be the largest value
with respect to $s_{13}^2$.
 Then,
the inequality results in
\begin{eqnarray}
 \mbb^\mi
  \le \left(m_1 c_{12}^2 + m_2 s_{12}^2 \right) c_\CH^2
        + m_3 s_\CH^2
\label{beta5}
\end{eqnarray}
for the normal hierarchy,
and
\begin{eqnarray}
 \mbb^\mi
  \le m_1 c_{12}^2 + m_2 s_{12}^2
\label{beta6}
\end{eqnarray}
for the inverted hierarchy.

 Constraints (\ref{beta3}), (\ref{beta5}) and (\ref{beta6})
determine an allowed region
on the plane of a neutrino mass versus the mixing angle.
 In this talk, we use the \mbox{$m_l$ - $\cos{2\theta_{12}}$} plane,
where $m_l$ denotes the lightest neutrino mass for each hierarchy.

\section{Discussion}
\begin{figure}[t]
\begin{center}
\hspace*{-30mm}
\includegraphics[scale=0.3]{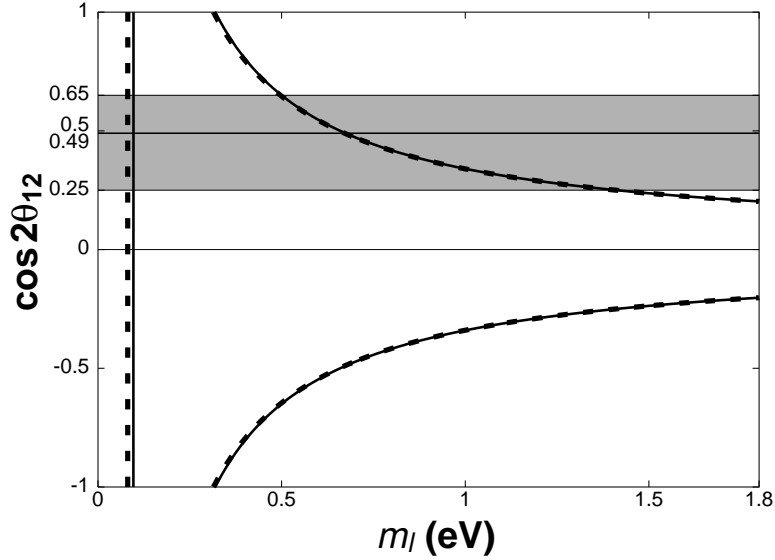}
\end{center}
\caption{
 The bounds (\ref{beta3}), (\ref{beta5}) and (\ref{beta6})
are shown for $0.1\,\eV \lesssim \mbb \lesssim 0.3\,\eV$.
 The solid (dashed) lines are for the normal (inverted) hierarchy.
 The inside of those bounds are allowed.
 The LMA region is superimposed with shadow.
}
\label{case1}
\end{figure}
 In this section, we analyze the constraints
obtained in the previous section.
 Two example cases of experimental results are considered below;
the case 1 is $0.1\,\eV \le \mbb \le 0.3\,\eV$
which is within the region of the KDHK result,
and the case 2 is $0.01\,\eV \le \mbb \le 0.03\,\eV$
which is outside of the region of the KDHK result.
 The LMA solution of the solar neutrino problem,
which is only one allowed at \mbox{99\%~C.L.}~\cite{SNO},
is considered mainly.
 Therefore, the mass square difference are fixed here after as
$|\Delta m^2_{12}| = 5.0\times10^{-5}\,\eV^2$ and
$|\Delta m^2_{23}| = 3.0\times10^{-3}\,\eV^2$.

\subsection{Case 1 : $0.1\,\eV \le \mbb \le 0.3\,\eV$}
 The bounds for this case are presented in Fig.~\ref{case1}.
 It is remarkable that
the bounds (\ref{beta5}) and (\ref{beta6}),
which are obtained with $\mbb^\mi$, are almost vertical
because of small $\Delta m_{12}^2$.
 The lines cross the horizontal axis at $m_l \simeq \mbb^\mi$
for not very small $\mbb^\mi$
(See the next subsection for very small $\mbb^\mi$).
 Therefore, $\mbb^\mi$ is approximately regarded
as the lower bound on $m_l$:$\mbb^\mi \lesssim m_l$.

 On the other hand,
since there are asymptotes $\cos{2\theta_{12}} = \pm t_\CH^2$
for the bounds (\ref{beta3}),
no upper bound on $m_l$ exists for $|\cos{2\theta_{12}}| \le t_\CH^2$.
 The LMA solution is fortunately outside of the region.
 Note that the bounds (\ref{beta3}) for the normal and inverted hierarchy
are very similar to each other.
 It means that the degenerate mass approximation $m_i \simeq m_\nu$
is very good in this case.
 In this approximation, the constraint (\ref{beta3}) becomes
\begin{equation}
 \mbb^\ma \ge m_\nu
               \left(
                c_\CH^2 \left| \cos{2\theta_{12}} \right| - s_\CH^2
	       \right) .
\end{equation}
 We see that $m_\nu/\mbb^\ma$ is a good parameter.
 In Fig.~\ref{degenerate}, the bounds (\ref{beta3}) are presented
on the $m_l/\mbb^\ma$ - $\cos{2\theta_{12}}$ plane
for $\mbb^\ma = 0.1\,\eV$\@.
 The similarity between the bounds for two hierarchies
means the goodness of the degenerate mass approximation
for $\mbb^\ma \simeq 0.1\,\eV$ and also for larger values of $\mbb^\ma$.
 Thus, the same bounds as those in Fig.~\ref{degenerate} can be used
for $\mbb^\ma \gtrsim 0.1\,\eV$.
 For example,
the simple constraint $m_l \le 2.2\times\mbb^\ma$ is read off in the figure
for the LMA best fit value $\cos{2\theta_{12}} = 0.49$,
which corresponds to $\tan^2\theta_{12} = 0.34$
given by the second article in~\cite{SNO}.

\subsection{Case 2 : $0.01\,\eV \le \mbb \le 0.03\,\eV$}
\begin{figure}[t]
\begin{center}
\hspace*{-30mm}
\includegraphics[scale=0.3]{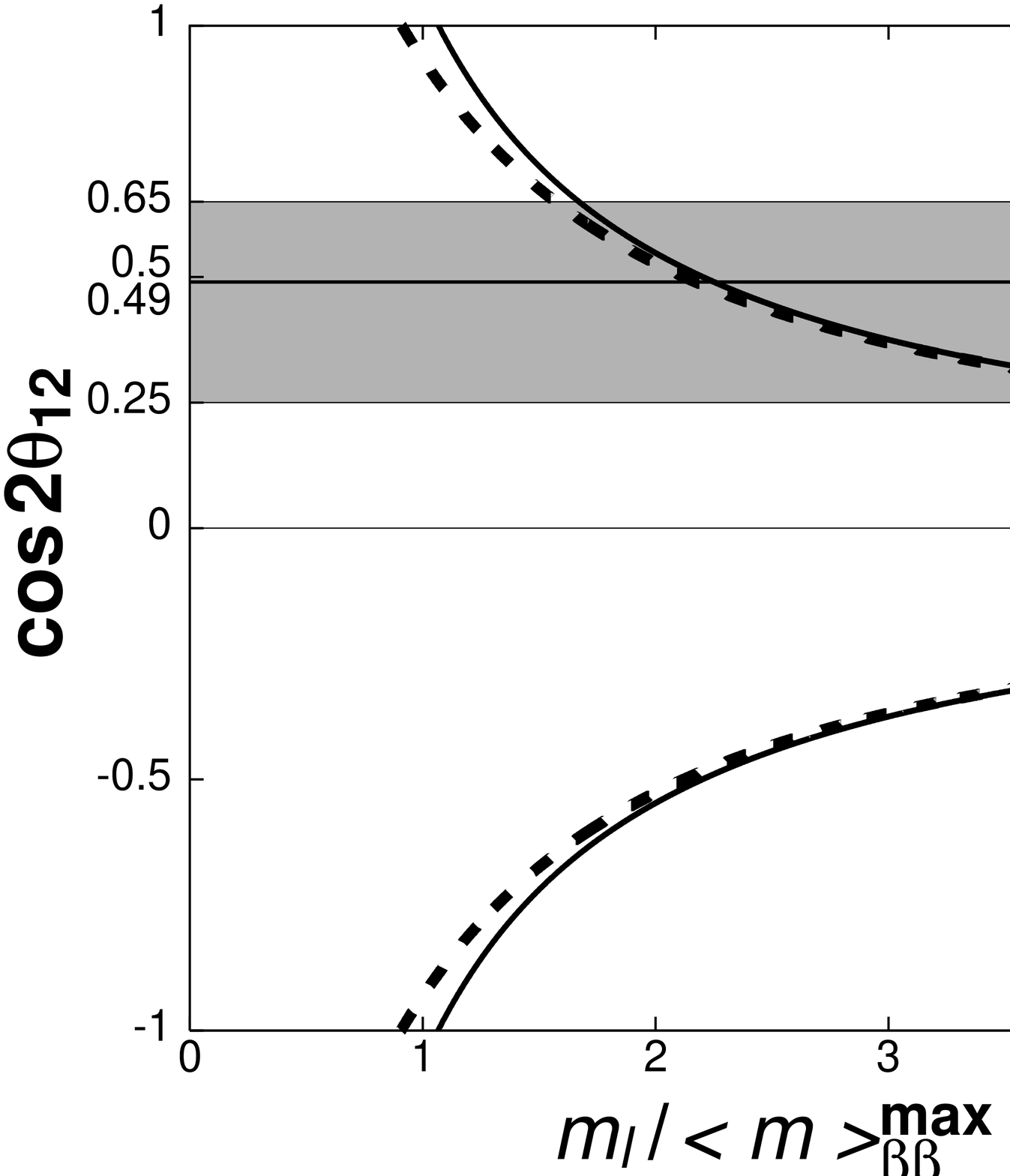}
\end{center}
\caption{
 The bounds (\ref{beta3}) are shown for $\mbb^\ma = 0.1\,\eV$.
 The solid (dashed) lines are for the normal (inverted) hierarchy.
 The inside of those bounds are allowed.
 The LMA region is superimposed with shadow.
}
\label{degenerate}
\end{figure}
\begin{figure}[t]
\begin{center}
\hspace*{-30mm}
\includegraphics[scale=0.3]{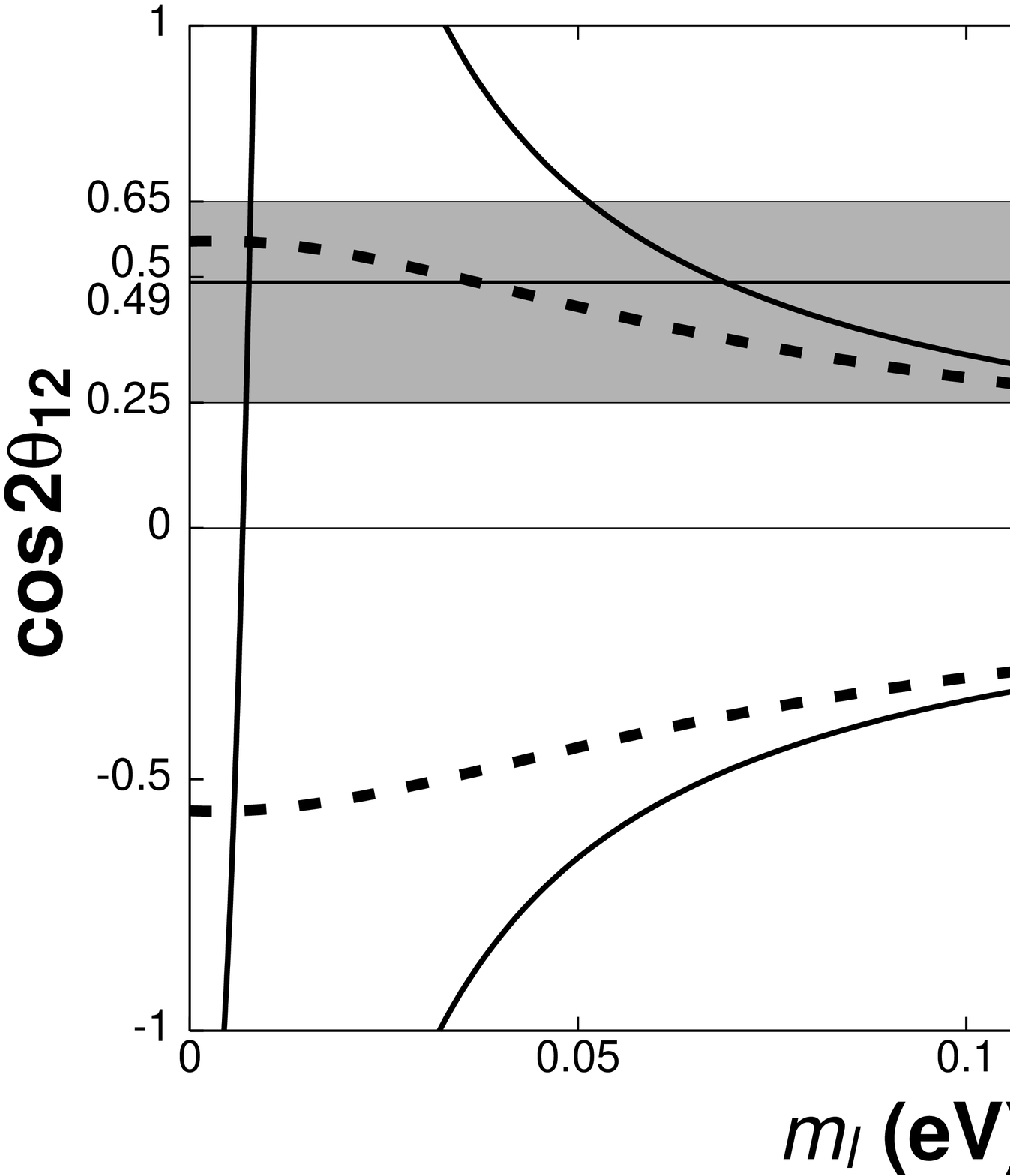}
\end{center}
\caption{
 The bounds (\ref{beta3}), (\ref{beta5}) and (\ref{beta6})
are shown for $0.01\,\eV \lesssim \mbb \lesssim 0.03\,\eV$.
 The solid (dashed) lines are for the normal (inverted) hierarchy.
 The inside of those bounds are allowed.
 The LMA region is superimposed with shadow.
}
\label{case2}
\end{figure}

 The bounds for this case are presented in Fig.~\ref{case2}.
 It is clear that the degenerate mass approximation is no longer good
because the relevant energy scale is smaller than
the atmospheric one $\sqrt{\Delta m^2_{23}} \simeq 0.05\,\eV$.
 The bounds for the normal and inverted hierarchy differ significantly.
 One of the important point is the disappearance of the bound
for the inverted hierarchy by (\ref{beta6}).
 It means that the constraint (\ref{beta6}) is satisfied
even for $m_l = 0$.
 It is also possible for the normal hierarchy with smaller $\mbb^\mi$.
 There are the smallest $\mbb^\mi$ needed for the existence of the bound
by (\ref{beta5}) and that by (\ref{beta6}).
 The values are extracted approximately from the RHS
of (\ref{beta5}) and (\ref{beta6}) with $m_l = \cos{2\theta_{12}} = 0$ as
\begin{equation}
 \frac{1}{\,2\,}\,c_\CH^2 \sqrt{\Delta m^2_{12}}
 + s_\CH^2 \sqrt{\Delta m^2_{23}}
 \simeq 0.005\,\eV \sim \sqrt{\Delta m^2_{12}}
\label{normal}
\end{equation}
for the normal hierarchy,
and
\begin{equation}
 \frac{1}{\,2\,}\,c_\CH^2
  \left(
   \sqrt{\Delta m^2_{23} - \Delta m^2_{12}} + \sqrt{\Delta m^2_{23}}
  \right)
 \simeq 0.053\,\eV \simeq \sqrt{\Delta m^2_{23}}
\label{inverted}
\end{equation}
for the inverted one.
 In the case 2, $\mbb^\mi = 0.01\,\eV$ is smaller than 0.053\,eV,
and that is why there is no lower bound on $m_l$ for the inverted hierarchy
in Fig.~\ref{case2}.

 Another important point of Fig.~\ref{case2} is that
the bound (\ref{beta3}) for the inverted hierarchy
crosses the vertical axis at $\cos{2\theta_{12}} = 0.57$.
 It means that
a small $\mbb$ excludes large values of $\cos{2\theta_{12}}$
for the hierarchy.
 Conversely, a $\cos{2\theta_{12}}$ larger than $t_\CH^2$
gives a theoretical minimum of $\mbb$\cite{mbbmin}.

 Finally, let us extract the bound on $m_l$ from the KDHK result.
 Since $\mbb^\ma = 0.84\,\eV$ is large enough,
we can use Fig.~\ref{degenerate}.
 Therefore, the upper bound on $m_l$ is extracted as
\begin{eqnarray}
 m_l \lesssim 2.2\times\mbb^\ma = 1.85\,\eV
\end{eqnarray}
for the LMA best fit parameters.
 On the other hand,
$\mbb^\mi = 0.05\,\eV$ is larger than (\ref{normal})
but smaller than (\ref{inverted}).
 Thus,
although there is the lower bound on $m_l$ for the normal hierarchy,
not for the inverted one.
 Roughly, $\mbb^\mi$ is the lower bound on $m_l$ for the normal hierarchy.
 By combining those results, we obtain
\begin{eqnarray}
 0.05\,\eV (0\,\eV) \lesssim m_l \lesssim 1.85\,\eV
\end{eqnarray}
for the normal (inverted) hierarchy
with the LMA best fit parameters.
 If $\mbb^\mi$ becomes a little larger,
$0\nu\beta\beta$ observations can give
the first exclusion of $m_l = 0$ for both hierarchies.

\section{Conclusions}
 Allowed regions on the plane of
a neutrino mass versus the solar mixing angle $\theta_{12}$
were obtained by using $\mbb^\ma$ ($\mbb^\mi$),
which is an experimental upper (lower) bound on
the observable $\mbb$ of double beta decay experiments.
 For given $\theta_{12}$,
these become constraints on a neutrino mass
such as the lightest mass $m_l$;
 Roughly, $\mbb^\mi \lesssim m_l \lesssim 2.2\times\mbb^\ma$
for the LMA best fit parameters.

 It became clear that
the condition $|\cos{2\theta_{12}}| > t_\CH^2 \simeq 0.03$
was necessary for the upper bound on $m_l$ to exist.
 On the other hand,
the condition $\mbb^\mi \gtrsim 0.005\,\eV (0.053\,\eV)$
needs to be satisfied
for the normal (inverted) hierarchy,
for the lower bound on $m_l$ to exist.

 For example,
$0.05\,\eV \le \mbb \le 0.84\,\eV$
gives $0.05\,\eV (0\,\eV) \lesssim m_l \lesssim 1.85\,\eV$
for the normal (inverted) hierarchy.

\section*{Acknowledgments}
 I would like to thank Prof.\ H.V.~Klapdor-Kleingrothaus
and other organizers of the conference
for the invitation and hospitality at Oulu.

\end{document}